\documentclass[preprint,showpacs,preprintnumbers,amsmath,amssymb]{revtex4}
\usepackage{graphicx}
\usepackage{color}
\usepackage{dcolumn}
\usepackage{bm}
\usepackage{mathrsfs}
\usepackage[T1]{fontenc}
\usepackage{amsmath,amssymb,graphics,epsfig,subfigure}

\begin{document}

\title{Implementing black hole as efficient power plant}

\author{Shao-Wen Wei \footnote{weishw@lzu.edu.cn},
        Yu-Xiao Liu\footnote{liuyx@lzu.edu.cn}}
\affiliation{Institute of Theoretical Physics $\&$ Research Center of Gravitation, Lanzhou University, Lanzhou 730000, People's Republic of China}

\begin{abstract}
Treating the black hole molecules as working substance and considering its phase structure, we study the black hole heat engine by a charged anti-de Sitter black hole. In the reduced temperature-entropy chart, it is found that the work, heat, and efficiency of the engine are free of the black hole charge. Applying the Rankine cycle with or without a back pressure mechanism to the black hole heat engine, the compact formula for the efficiency is obtained. And the heat, work and efficiency are worked out. The result shows that the black hole engine working along the Rankine cycle with a back pressure mechanism has a higher efficiency. This provides a novel and efficient mechanism to produce the useful mechanical work, and such black hole heat engine may act as a possible energy source for the high energy astrophysical phenomena near the black hole.
\end{abstract}


\maketitle

\section{Introduction}
\label{secIntroduction}

Black hole has been a fascinating object since general relativity predicted its existence. Compared with other stars, a black hole has sufficient density and a huge amount of energy. Near a black hole, there are many high energy astrophysical phenomena related to the energy flow, such as the power jet and black holes merging. Thus extracting energy from or using black hole is an interesting and outstanding challenge subject.

As early as 1969, Penrose~\cite{Penrose} proposed that a particle, at rest at infinity, arrives the ergosphere of a Kerr black hole and then decays into two photons, one of which with negative energy crosses the event horizon, while other one with positive energy escapes to infinity. This process provides us a mechanism to extract the rotational energy of the black hole, and the maximum efficiency for a single particle via this process is 121\%~\cite{Wald}.

Several years ago, Banados, Silk, and West~\cite{Banados} showed that Kerr black holes can act as high energy particle accelerators and the center of mass energies of two colliding particles could reach the Planck scale when some fine tunings are satisfied. This mechanism was also extended to the collisional Penrose process that two particles collide inside the ergosphere of a rotating black hole and produce two particles, one of which falls into the black hole while other one escapes to infinity. The net efficiency of the process can approach to 130\%~\cite{Bejger} for a single particle, or to a higher efficiency~\cite{Schnittman,Berti}. Replacing the two collided particles with two black holes, there will be the black holes merging, which is a violent astrophysical phenomenon in our Universe. During the collision, a large amount of energy is released through the gravitational waves, which were recently directly observed by the LIGO scientific Collaboration~\cite{Abbott}. Black hole bomb~\cite{Press,Cardoso} is also another interesting mechanism to extract energy from black holes due to the superradiant instability.

Moreover, energy extraction can also be achieved with a thermodynamic method. Dolan~\cite{Dolan} pointed out that, in an isentropic and isobaric process, the anti-de Sitter (AdS) black hole can yield mechanical work by decreasing its angular momentum. And an extremal electrically neutral or charged rotating black hole can have an efficiency of up to 52\% or 75\%. In Ref.~\cite{Bravetti}, Bravetti, Gruber, and Lopez-Monsalvo considered thermodynamic optimization of Penrose process using the finite-time thermodynamics.

Most recently, Johnson~\cite{Johnson} operated a simple holographic black hole heat engine along a thermodynamic cycle to extract the useful mechanical work. The exact efficiency formula was also obtained entirely in terms of the black hole mass~\cite{Johnson2}. This provides a novel thermodynamic way to extract energy using the black hole, as well as a possible application of AdS/CFT correspondence. A remarkable character of the heat engine is that it exchanges the heat and work with its surroundings. For a black hole engine, we refer the surroundings to the accretion disk around the black hole. To some extent, we can think that they can arrive the steady state equilibrium. Thus the accretion disk could play the role of the heat source of sink for the black hole engine. If the temperature of the accretion disk gets a closed cycle, then the black hole will also complete a thermodynamic cycle. Then a net work may be produced and therefore a black hole heat engine will be fulfilled.

Importantly, for the implementation of a black hole heat engine, we argue that the thermodynamic property and phase transition of the working substance should be included in for two reasons: i) a cycle is crucially built by utilizing the thermodynamic property of the working substance. For example, water working in a steam power plant will undergo liquid-vapour phase transition during each cycle. ii) Raising the engine efficiency through reducing the low-temperature heat sources will unavoidably encounter the phase transition of the substance. For a black hole engine, we here suggest that the working substance is the fluid constituted with the virtual black hole molecules~\cite{Wei}, which carry the degrees of freedom of black hole entropy. Thus, the aim of this paper is to investigate the black hole heat engine implemented with a practical thermodynamic cycle when the thermodynamic property and the phase transition of the black hole molecules is included in.

The paper is organized as follows. In Sec. \ref{Classification}, we give a brief review of the phase diagram for the charged AdS black hole. Then we, respectively, construct and study the Carnot and Rankine cycles in Secs. \ref{carnotcycle} and \ref{Rankine}. The summary and remarks follow in Sec. \ref{Summary}.

\section{Phase diagram}
\label{Classification}

\begin{figure}
\includegraphics[width=8cm]{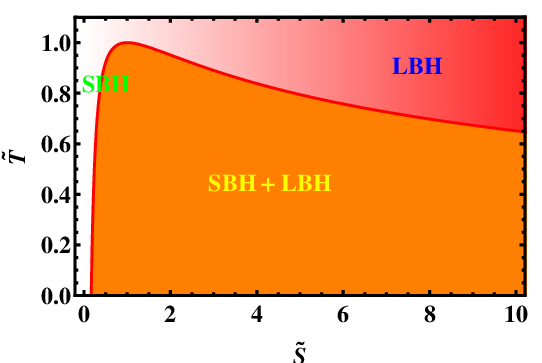}
\caption{Phase diagram in the reduced $\tilde{T}$-$\tilde{S}$ chart. The red thick line denotes the coexistence curve. Below the curve is the region of coexistence phase of small and large black holes. Left and right regions above the curve are the small and large black hole phases, respectively. Here we call the left part and right part of the coexistence curve as the saturated small and large black hole curves, respectively.} \label{WenShang}
\end{figure}

Let us first proceed with the thermodynamic property of the working substance. In the extended phase space, the cosmological constant is treated as a pressure,
\begin{eqnarray}
 P=-\frac{\Lambda}{8\pi},
\end{eqnarray}
and the black hole mass is interpreted as the enthalpy~\cite{Kastor}. In this approach, the interpretation of many other thermodynamic quantities is modified~\cite{Dolan2}. For the four-dimensional charged AdS black hole, the line element is
\begin{eqnarray}
 ds^{2}&=&-f(r)dt^{2}+\frac{dr^{2}}{f(r)}+r^{2}(d\theta^{2}+\sin^{2}\theta d\phi^{2}),\label{metric}\\
 F&=&dA,\quad A=-\frac{q}{r}dt.
\end{eqnarray}
where the metric function is given by
\begin{eqnarray}
 f(r)=1-\frac{2M}{r}+\frac{q^{2}}{r^{2}}+\frac{r^{2}}{l^{2}}.
\end{eqnarray}
The parameters $M$ and $q$ relate to the black hole mass and charge. Using the `Euclidean trick', the black hole temperature can be obtained
\begin{eqnarray}
 T=\frac{1}{4\pi r_{h}}(1+\frac{3r_{h}^{2}}{l^{2}}-\frac{q^{2}}{r_{h}^{2}}),
 \label{temp}
\end{eqnarray}
with $r_{h}$ being the radius of the black hole event horizon. The black hole entropy is given by
\begin{eqnarray}
 S=\frac{A}{4}=\pi r_{h}^{2}.
\end{eqnarray}
Then we can rewrite the temperature (\ref{temp}) as
\begin{eqnarray}
 T=\frac{-\pi q^{2}+S+8P S^{2}}{4\sqrt{\pi} S^{3/2}}.
\end{eqnarray}
And the enthalpy of the black hole system is~\cite{Kastor}
\begin{eqnarray}
 H\equiv M=\frac{3\pi q^{2}+3S+8PS^{2}}{6\sqrt{\pi S}}.
\end{eqnarray}
After the study of this state equation, it is found that there exists a small-large black hole phase transition of van der Waals type~\cite{Kubiznak}. By solving $(\partial_{S}T)_{P,q}=(\partial_{S,S}T)_{P,q}=0$, one can get the critical point~\cite{Kubiznak}
\begin{eqnarray}
 P_{c}=\frac{1}{96\pi q^{2}},\quad
 T_{c}=\frac{\sqrt{6}}{18\pi q},\quad
 S_{c}=6\pi q^{2}.
\end{eqnarray}
In the reduced parameter space, the state equation describing this phase transition can be expressed in the following form
\begin{equation}
 \tilde{T}=\frac{1}{8\tilde{S}^{3/2}}\left(3\tilde{P}\tilde{S}^{2}+6\tilde{S}-1\right),
\end{equation}
where the reduced physical quantities are defined as $\tilde{A}=A/A_{c}$ with $A_{c}$ being the critical value. The phase diagram is shown in the reduced temperature-entropy chart in Fig. \ref{WenShang}. The red thick line denotes the coexistence curve, which has the analytical form
\begin{equation}
 \tilde{T}^{2}=1+\cos\left(3\arccos(\frac{2+\tilde{S}-\sqrt{3+6\tilde{S}}}{2\tilde{S}})\right),
\end{equation}
where $\tilde{S}\geq1/6$. Below this curve is the coexistence region of small and large black holes. Left and right regions above the coexistence curve are the small and large black hole phases, respectively. Note that in the $P$-$T$ chart the coexistence curve has been given in Ref.~\cite{Spallucci}. On the other hand, when crossing the coexistence curve, the microscopic structure of the black hole will be greatly changed~\cite{Wei}. One of the features in the reduced parameter space is that all the thermodynamic quantities are free of the black hole charge.

\section{Carnot cycle}
\label{carnotcycle}

After achieving the property of the working substance, one can build a cycle for a heat engine. At first, let us turn to the thermodynamic law during a physical process. For a closed system and a steady-state flow system undergoing an infinitesimal thermodynamic process, we have, respectively,
\begin{eqnarray}
 dU&=&TdS-PdV,\\
 dH&=&TdS+VdP.\label{fistlaw}
\end{eqnarray}
The heat change is $\text{\dj}Q=TdS$ with `$\text{\dj}$' denoting $Q$ is not a state function. While the work change is \dj$W=-PdV$ for the closed system, and $\text{\dj}W=VdP$ for the steady-state flow system. Thus one arrives at the first law $dU(dH)=\text{\dj}Q+\text{\dj}W$ for the two systems. It is worthwhile to point out that $dU$ is the amount of energy stored in the system, while $dH$ measures the energy stored by the working substance rather than the system.

Considering a reversible thermodynamic process, i.e., the system follows a path transferring state A to state B, the work and heat can be obtained with integrating the first law from A to B. For simplicity, we can cast the process into the $P$-$V$ chart, in which the area constructed with the path and $V$-axis is equal to the negative work for the closed system, and the area constructed with the path and $P$-axis is equal to the work for the steady-state flow system. In addition, the $T$-$S$ chart is complementary to the $P$-$V$ chart. For the two systems, the heat can be both measured with the area under the path. Therefore we can obtain the work or heat through measuring the areas by casting the process into the $P$-$V$ chart or $T$-$S$ chart.

If a device operates in a cyclical manner along a closed thermodynamic process and exchanges only heat and work with its surroundings, we call it a heat engine. And we require that no working substance is lost or gained during the cycle. For a simple heat engine, it shares two characteristics: a heat source at high temperature and a heat sink at lower temperature. At high temperature, the engine absorbs the heat $Q_{1}$ (positive) from the high temperature source, and $Q_{2}$ (negative) from the heat sink. During one cycle, we have $\Delta U=0$ or $\Delta H=0$, which leads to
\begin{eqnarray}
 Q_{1}+Q_{2}+W=0,
\end{eqnarray}
where $W$ (negative) measures the work done by the surroundings to the engine during one cycle. The closed and steady-state flow systems share the same expression, so we will not distinguish them. Another important characteristic parameter of a heat engine is its efficiency, which is defined as the ratio of work output from the machine and heat input at high temperature,
\begin{eqnarray}
 \eta=-\frac{W}{Q_{1}}=1+\frac{Q_{2}}{Q_{1}}.
\end{eqnarray}
Since $Q_{2}<0$, we have $\eta<1$. Explicitly, in order to obtain the efficiency, one needs to calculate $Q_{1}$ and $Q_{2}$. Thus we can cast the cycle into the $T$-$S$ chart only, and then the efficiency can be obtained by measuring the areas under the paths.

\begin{figure}
\center{\subfigure[Carnot cycle]{\label{pCarnot}
\includegraphics[width=6cm]{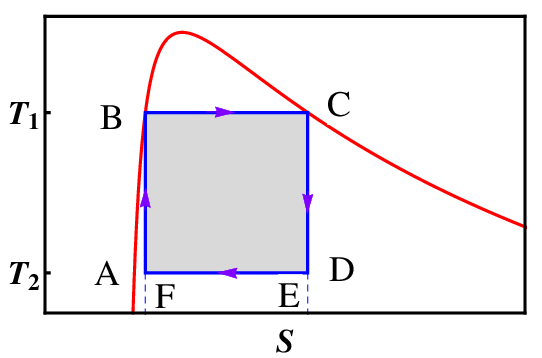}}
\subfigure[Rankine cycle]{\label{pRankine}
\includegraphics[width=6cm]{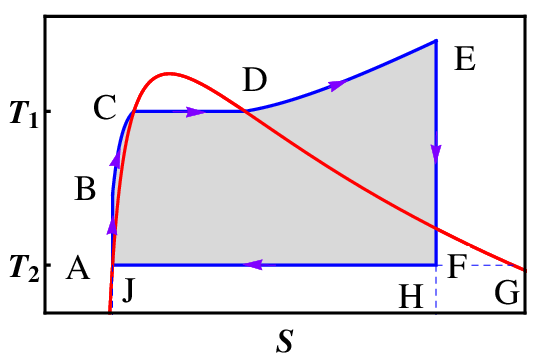}}}
\caption{Sketch pictures of thermodynamic cycle for heat engines. Thick red lines denote the coexistence curve. (a) Carnot cycle and (b) Rankine cycle.}
\end{figure}

Next, we will turn to the Carnot engine, which is a particularly simple design conceived by Carnot in 1824. The Carnot cycle contains four reversible steps, i.e., two isothermal steps and two adiabatic steps. In general, the Carnot cycle is described in the $P$-$V$ chart, which appears as an irregular loop. However, when casting this four steps in the $T$-$S$ chart, one will find that this cycle is a regular rectangle, and its area can be quite easily calculated. We sketch the cycle in the $T$-$S$ chart in Fig. \ref{pCarnot}. Steps A-B and C-D described by the vertical lines are adiabatic. And steps B-C and D-A described by the horizontal lines are two isothermal steps. For a steady-state flow Carnot engine, steps D-A and A-B are implemented with compressor, and steps B-C and C-D with expansion turbine. Then the efficiency for the Carnot engine will be
\begin{eqnarray}
 \eta=1+\frac{Q_{2}}{Q_{1}}=1-\frac{\text{area(ADEFA)}}{\text{area(BCEFB)}}=1-\frac{T_{2}}{T_{1}}.
\end{eqnarray}
The useful work done by the engine in one cycle is just the negative of the area enclosed by the cycle.

Before applying this cycle to a black hole engine, one needs to be familiar with the thermodynamic property of the working substance. As stated above, the black hole system exhibits a small-large black hole phase transition. And in the reduced parameter space, the phase structure is found to be charge-free. So here we would like to extend the study to the reduced $\tilde{T}$-$\tilde{S}$ chart. The reduced heat and work can be defined as
\begin{eqnarray}
 \tilde{Q}&=&Q/T_{c}S_{c}=\int \tilde{T}d\tilde{S},\\
 \tilde{W}&=&W/T_{c}S_{c}.
\end{eqnarray}
After doing this, $\tilde{Q}$ and $\tilde{W}$ will be free of the black hole charge. And the efficiency for the Carnot cycle is given by $\eta=1-\tilde{T}_{2}/\tilde{T}_{1}$ .

\begin{figure}
\includegraphics[width=8cm]{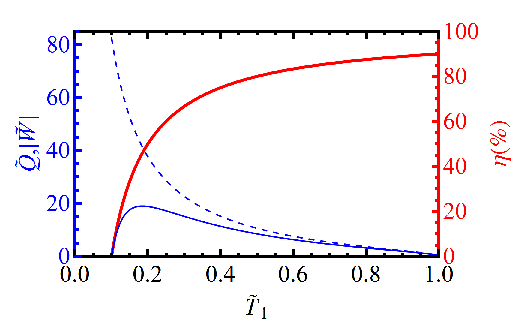}
\caption{Carnot engine with $\tilde{T}_{2}$=0.1. The work and heat for the engine are plotted in thin solid blue line and thin dashed blue line (left $y$-axis), respectively. The engine efficiency is plotted in thick solid red line (right $y$-axis).}\label{pnt}
\end{figure}

For simplicity, we require that the working substance of the black hole heat engine is always in the coexistence phase, which means the cycle should locate under the coexistence curve in the $\tilde{T}$-$\tilde{S}$ chart showed in Fig. \ref{pCarnot}. In this case, we consider a maximal Carnot cycle (points B and C lie on the coexistence curve), which has the maximal work among all the cycles operating between the same two temperatures. Here we fix the low temperature $\tilde{T}_{2}=0.1$ and let the high temperature $\tilde{T}_{1}$ vary. The reduced heat $\tilde{Q}$, work $\tilde{W}$, and efficiency $\eta$ are plotted against $\tilde{T}_{1}$ in Fig. \ref{pnt}. From it, we can see that the reduced work firstly starts at $\tilde{T}_{1}=0.1$, then approaches its maximum at $\tilde{T}_{1}=0.1849$, and finally decreases to zero at $\tilde{T}_{1}=1$, where the critical temperature is approached. While the reduced heat behaves very differently from the reduced work. It monotonously decreases with $\tilde{T}_{1}$, and approaches to zero at the critical temperature. The efficiency $\eta$ increases with $\tilde{T}_{1}$, and arrives 0.9 at $\tilde{T}_{1}=1$. This states that with fixed low temperature, raising the high temperature can give a high efficiency for the engine.

\section{Rankine cycle}
\label{Rankine}

Although the reversible heat Carnot engine has the highest efficiency among all heat engines operating between the same two temperatures, it is difficult to implement mechanically. A practical and extensively used heat engine is the steam power plant, which can be modeled with a Rankine cycle showed in Fig. \ref{pRankine}. During such cycle, the working fluid will undergo a phase transition. For the black hole, this cycle is completely new and is worth to pursue. Like the steam power plant, such Rankine cycle may provide a practical energy source for the high energy astrophysical phenomena near the black hole. Here we give a simple introduction for the Rankine cycle. The black hole molecules, which act as the working fluid, start at state A and approach to state B in an adiabatic process. Next they follow one isobaric line from state B to state E. During this process, these molecules undergo a small-large black hole phase transition from state C to state D with a constant temperature. Then the molecules perform a useful work and reduce the temperature from state E to state F such that the molecules change from the large black hole phase to a coexistence phase of small and large black holes. Finally, these molecules return to state A with shrinking their specific volume.

For the steam power plant, A-B process is implemented with a feed-pump, B-E process with a boiler of the steam generator and a superheater, E-F process with a turbine, and F-A process with a condenser. For the black hole engine, we may imagine that there exist four effective mechanisms to implement the four processes. For example, at first the black hole and its surroundings, the accretion disk, are in a state of equilibrium corresponding to a saturated small black hole state A. Then the temperature of the disk increases for some reasons like the friction of the disk gas molecules. Subsequently, this leads to a little increase of the black hole temperature. Since such process is very short, the black hole may keep its size unchanged. This is just related to an adiabatic A-B process. Next the black hole gets hotter and hotter along the B-E process with the increase of the disk temperature. During this process, the black hole experiences a small-large black hole phase transition. When the black hole achieves the state E, this black hole system may encounter a violent physical phenomena, such as a sudden jet, which will result in a sudden drop of the temperature both for the black hole and its accretion disk. During this rapid process, the black hole may also keep its size unchanged, and that process is just described by the E-F process. After that, the black hole will gradually approach to state A with decreasing its size while maintaining a constat temperature as that of the accretion disk.

\begin{figure}
\includegraphics[width=8cm]{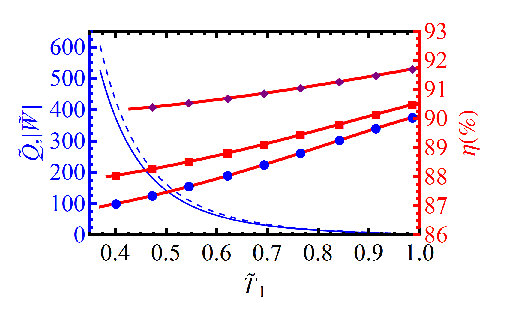}
\caption{Rankine cycle with $\tilde{T}_{2}$=0.1. The reduced work and heat for the engine are plotted in thin solid blue line and thin dashed blue line (left $y$-axis), respectively. The engine efficiency is plotted in thick solid red lines (right $y$-axis) for $\tilde{T}_{E}$=1.1, 1.2, and 1.5 from bottom to top.}\label{pn1a}
\end{figure}

Next, we would like to turn to the calculation of the efficiency for the Rankine cycle. During one cycle, the black hole engine only absorbs heat along the B-E process and releases heat in the H-J process. According to the analysis in Sect. \ref{carnotcycle}, the heat can be measured with the area under the process showed in the $T$-$S$ chart,
\begin{eqnarray}
 Q_{1}&=&\text{area(JBCDEHJ)},\\
 Q_{2}&=&-\text{area(JAFHJ)}.
\end{eqnarray}
The absolute value of the net work output from the E-F process is
\begin{eqnarray}
 |W|=Q_{1}+Q_{2}=\text{area(ABCDEFA)}.
\end{eqnarray}
Thus, the efficiency for this Rankine cycle can be calculated as
\begin{eqnarray}
 \eta=-\frac{W}{Q_{1}}=1+\frac{Q_{2}}{Q_{1}}=1-\frac{\text{area(AFHJA)}}{\text{area(JBCDEHJ)}}.
\end{eqnarray}
In the following, we would like to show how to measure the areas for these thermodynamic processes. With the help of Fig. \ref{pRankine}, the heats $Q_{1}$ and $Q_{2}$ can be expressed as
\begin{eqnarray}
 Q_{1}&=&Q_{BE}=Q_{BC}+Q_{CD}+Q_{DE},\\
 Q_{2}&=&Q_{FA}.
\end{eqnarray}
Since B-C and E-F processes are adiabatic, there is no heat absorbed or released. Processes B-C and D-E are isobaric processes with the same pressure, so the heats are
\begin{eqnarray}
 Q_{BC}&=&\int_{B}^{C}C_{P}dT
        =\int_{B}^{C} T dS,\label{bc}\\
 Q_{DE}&=&\int_{D}^{E}C_{P}dT
        =\int_{D}^{E} T dS,\label{de}
\end{eqnarray}
where $C_{P}$ is the heat capacity of the black hole with fixed pressure. And from this equation, we are clear that the heat is just the area under the process. On the other hand, for the isothermal process, the heats are much easier to get, i.e.,
\begin{eqnarray}
 Q_{CD}&=&\int_{B}^{C} T_{1} dS=T_{1}(S_{D}-S_{C}),\\
 Q_{FA}&=&\int_{F}^{A} T_{2} dS=T_{2}(S_{A}-S_{F}).
\end{eqnarray}
In our approach, one has $S_{A}<S_{F}$, so $Q_{FA}$ is negative. The temperature $T_{1}$ is a phase transition temperature. By using the equal area law, we have
\begin{eqnarray}
 T_{1}(S_{D}-S_{C})=\int_{C}^{D} T dS.
\end{eqnarray}
Combining with Eqs. (\ref{bc}) and (\ref{de}), the heat absorbed in the processes B-C, C-D and D-E can be expressed in a compact form
\begin{eqnarray}
 Q_{BE}=\int_{B}^{E} T dS.\label{be}
\end{eqnarray}
For the isobaric process, the first law (\ref{fistlaw}) will be of the following form
\begin{eqnarray}
 dH=TdS.
\end{eqnarray}
Performing a integral, one easily gets
\begin{eqnarray}
 H=\int TdS.
\end{eqnarray}
Comparing with (\ref{de}), we finally arrive
\begin{eqnarray}
 Q_{BE}=H_{E}-H_{B}.
\end{eqnarray}
Utilizing this result, the efficiency for the Rankine cycle can be expressed as
\begin{eqnarray}
 \eta=1-\frac{T_{2}(S_{F}-S_{A})}{H_{E}-H_{B}}.
\end{eqnarray}
The efficiency will be calculated by employing this formula. Interestingly, we can also express the efficiency in the reduced parameter space
\begin{eqnarray}
 \eta=1-\frac{\tilde{T}_{2}(\tilde{S}_{F}-\tilde{S}_{A})}{\tilde{H}_{E}-\tilde{H}_{B}}.
\end{eqnarray}
It is important to point out that the reduced enthalpy is defined as
\begin{eqnarray}
 \tilde{H}=\frac{H}{T_{c}S_{c}}.
\end{eqnarray}
We show the reduced heat and work in Fig. \ref{pn1a}. The low temperature is set to $\tilde{T}_{2}$=0.1, while $\tilde{T}_{1}$ freely varies. One thing worth to note is that during the cycle, state F should fall into the coexistence range, thus $\tilde{T}_{1}$ has a minimum value larger than $\tilde{T}_{2}$. $\tilde{Q}_{1}$ and $\tilde{W}$ are plotted with $\tilde{T}_{E}$=1.1. At the minimum temperature $\tilde{T}_{1}=0.37$, the reduced heat and work have maximal values, i.e., $\tilde{Q}_{1}=602$ and $\tilde{W}=524$. When increasing $\tilde{T}_{1}$ from its minimum, $\tilde{Q}_{1}$ and $\tilde{W}$ decrease, and approach to 3.68 and 3.32 at $\tilde{T}_{1}$=1.

The efficiency is described in Fig. \ref{pn1a} with thick red lines for $\tilde{T}_{E}$=1.1, 1.2, and 1.5, respectively, from bottom to top. It monotonously increases with $\tilde{T}_{1}$. And the higher the $\tilde{T}_{E}$, the higher the efficiency. For clarity, we present the work and efficiency for the Rankine cycle with $\tilde{T}_{2}$=0.1 and $\tilde{T}_{E}$=1.1 in Table \ref{tab0}. From it, we can see that low $\tilde{T}_{1}$ and high $\tilde{T}_{1}$ behave very differently. For example, the work $W$=376 and the efficiency $\eta$=87.07\% for low $\tilde{T}_{1}$=0.40. While the work will greatly decreases and efficiency increases for high $\tilde{T}_{1}$. For a comparison, $W$=4.75 and $\eta$=89.92\% for the case $\tilde{T}_{1}$=0.96.

In summary, for the Rankine cycle, we can clearly obtain two results: i) at low reduced temperature $\tilde{T}_{1}$, we have more useful work while low efficiency. ii) at high $\tilde{T}_{1}$, we get higher efficiency but less useful work.

\begin{table}[h]
\begin{center}
\begin{tabular}{ccccc}
  \hline\hline
  $\tilde{T}_{1}$ & 0.40/0.96 & 0.41/0.97 & 0.42/0.98 & 0.43/0.99\\\hline
  $\tilde{W}$ & 376/4.75 & 338/4.36 & 305/4.00 & 275/3.65\\ \hline
  $\eta$(\%) & 87.07/89.92 & 87.11/89.97 & 87.15/90.01 & 87.19/90.06\\\hline\hline
\end{tabular}
\caption{Work and efficiency for the Rankine cycle with $\tilde{T}_{2}$=0.1 and $\tilde{T}_{E}$=1.1.}\label{tab0}
\end{center}
\end{table}

In practice, one may expect more useful work and  higher efficiency for a heat engine, which seems to be very hard to realize in this case. A compromise is that we can operate the engine at a not high temperature $\tilde{T}_{1}$ and then find a way to raise the efficiency. As shown above, raising $\tilde{T}_{E}$ could provide us a way to a high efficiency. And for a steam power plant, this mechanism is implemented with a back pressure turbine. Its effect is to make the point F fall on the coexistence curve. For a black hole heat engine, this requires that the end state of the cooling process E-F is a saturated large black hole. In fact, there exists a highest temperature for $T_{E}$. When the black hole temperature approaches it, the energy release mechanism is triggered, and the black hole temperature rapidly decreases. Maybe such trigger mechanism is formulated by the property of the disk gas molecules or some other unforeseen techniques. Nevertheless, it is of a great theoretical interest to consider such thermodynamic cycle. For simplicity, we ignore the possible extra energy to realize this back pressure mechanism.

\begin{figure}
\includegraphics[width=8cm]{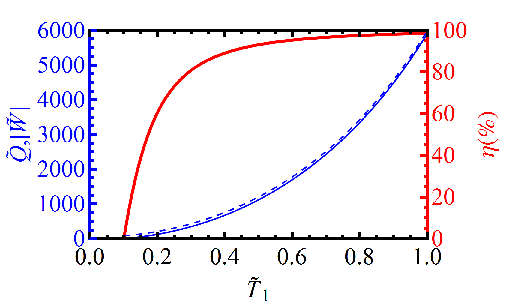}
\caption{Rankine cycle with back pressure mechanism and $\tilde{T}_{2}$=0.1. The reduced work and heat for the engine are plotted in thin solid blue line and thin dashed blue line (left $y$-axis), respectively. The engine efficiency is plotted in thick solid red line (right $y$-axis).}\label{bfrankine}
\end{figure}

\begin{figure}
\center{
\includegraphics[width=8cm]{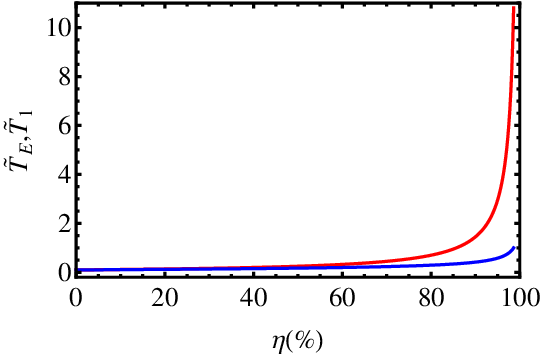}}
\caption{Temperatures $\tilde{T}_{E}$ (above) and $\tilde{T}_{1}$ (below) vs. efficiency for the BPR cycle with $\tilde{T}_{2}$=0.1.}\label{bT1TE}
\end{figure}

Here we refer this cycle as the back pressure Rankine (BPR) cycle. For the case, one has
\begin{eqnarray}
 T_{2}(S_{F}-S_{A})=H_{F}-H_{A}.
\end{eqnarray}
Thus its efficiency reads
\begin{eqnarray}
 \eta=1-\frac{H_{F}-H_{A}}{H_{E}-H_{B}},
\end{eqnarray}
or in the reduced parameter space
\begin{eqnarray}
 \eta=1-\frac{\tilde{H}_{F}-\tilde{H}_{A}}{\tilde{H}_{E}-\tilde{H}_{B}}.
\end{eqnarray}
The reduced heat, work, and efficiency are shown in Fig. \ref{bfrankine}. The low temperature is also set to $\tilde{T}_{2}$=0.1. From the figure, one can clearly see that the behavior of this BPR cycle is very different from the Rankine cycle without the back pressure mechanism. Both more work and higher efficiency can be achieved at the same time. Thus, the left thing is to rise the reduced temperature $\tilde{T}_{1}$ for a high efficiency. However, we should keep in mind that $\tilde{T}_{1}$ is not the highest temperature for the engine to operate, while $\tilde{T}_{E}$ is. Since the normal working of the heat engine is limited with the temperature, $\tilde{T}_{E}$ should be not too high. We clearly show the temperatures $\tilde{T}_{E}$ and $\tilde{T}_{1}$ as functions of the efficiency in Fig. \ref{bT1TE}. For low efficiency, $\tilde{T}_{E}$ almost meets $\tilde{T}_{1}$. However, when the efficiency increases up to 60\%, the difference between them will be obvious. When exceed that value, the temperature $\tilde{T}_{E}$ has a rapidly increase than $\tilde{T}_{1}$. More precisely, we list some numerical data in Table \ref{tab1}, which shows that in order to achieve the efficiency $\eta$, how high the temperatures $\tilde{T}_{1}$ and $\tilde{T}_{E}$ should be. For small $\eta$, the highest temperature $\tilde{T}_{E}$ of the heat engine is small. However if one wonder an efficiency larger than 90\%, $\tilde{T}_{E}$ will rapidly increase. For example, $\eta=96\%$ requires $\tilde{T}_{E}=3.70$, and $\eta=98\%$ requires $\tilde{T}_{E}=7.45$. Nevertheless, the BPR cycle still has a higher efficiency than the standard Rankine cycle.

\begin{table}[h]
\begin{center}
\begin{tabular}{cccccccc}
  \hline\hline
  $\eta$(\%)  & 20 &  60 & 80 & 90 & 93 & 96 & 98  \\\hline
  $\tilde{T}_{1}$ & 0.12 &  0.20 & 0.29 & 0.42 & 0.50 & 0.64 & 0.86\\ \hline
  $\tilde{T}_{E}$ & 0.14 &  0.32 & 0.70 & 1.45 & 2.09 & 3.70 & 7.45\\\hline\hline
\end{tabular}
\caption{Efficiency and temperatures for the BPR cycle with $\tilde{T}_{2}$=0.1.}\label{tab1}
\end{center}
\end{table}

\section{Summary}
\label{Summary}

In this paper, we applied the Rankine cycle to the black hole heat engine by analogy with the steam power plant. We first examined the phase structure for the black hole molecules working as substance in the engine. Then after making a comparison between the small-large black hole phase transition and the liquid-vapour phase transition of water, we operated the black hole engine along the Rankine cycle with which the steam power plant operates. In the reduced $\tilde{T}$-$\tilde{S}$ chart, the work, heat, and efficiency of the engine were found to be charge-free. Furthermore, our result shows that, when a back pressure mechanism is introduced for the black hole heat engine, the engine efficiency will approach its maximum. For example, the efficiency for $\tilde{T}_{2}=0.1$ reaches $98.62\%$, which is only slightly smaller than the bounded efficiency 99.07\% of the corresponding Carnot engine. More importantly, the engine also produces the maximal work at the same time. Thus this result may provide us a novel and efficient mechanism to produce the useful mechanical work with the black hole.

At last, we would like to make a few comments. Here we dealt with the charged AdS black hole heat engine operating along the Rankine cycle. During the cycle, the engine produces useful work to its surroundings. Of course, one can consider a reverse cycle, the refrigeration cycle for the black hole engine if the work is provided by its surroundings. Moreover, one can consider the heat engine by the rotating AdS black holes or black holes in higher-derivative gravity, where the phase diagram is similar to the water with a triple point being found. These may provide a novel way to transfer heat energy to mechanical work, and black hole heat engine may act as a possible energy source for the high energy astrophysical phenomena near the black holes. On the other hand, according to the AdS/CFT correspondence, finding out the holographic dual description of such thermodynamic cycle in the large $N$ field theory is also an interesting and important subject. It is worth to point out that in Ref.~\cite{Rosso}, the authors have expressed the heat and work of the heat engines by using the field theory terms. Further, understanding the CFTs via the property of the holographic heat engines is also very promising.

\section*{Acknowledgements}
This work was supported by the National Natural Science Foundation of China (Grants No. 11675064, No. 11522541, and No. 11375075), and the Fundamental Research Funds for the Central Universities (Grants Nos. lzujbky-2016-115).

\end{document}